\documentclass{article}
\usepackage{graphicx} 
\usepackage{amsmath}
\usepackage{lipsum}
\usepackage{amsfonts}
\usepackage{amssymb}
\usepackage{amsbsy}
\usepackage{mathrsfs}
\usepackage{amsthm}
\usepackage{graphicx}    
\usepackage{rotating}    
\usepackage{epsfig}
\usepackage{ulem}
\usepackage{color}       
\usepackage{xcolor}

\definecolor{gruen}{rgb}{0,0.625,0}     
\definecolor{rot}{rgb}{0.75,0,0}        
\definecolor{blau}{rgb}{0,0,0.75}       

\newcommand{\BEQ}{\begin{equation}}     
\newcommand{\BEA}{\begin{eqnarray}}
\newcommand{\BD}{\begin{displaymath}}
\newcommand{\EEQ}{\end{equation}}       
\newcommand{\EEA}{\end{eqnarray}}
\newcommand{\ED}{\end{displaymath}}
\newcommand{\D}{{\rm d}}                
\newcommand{\II}{{\rm i}}               
\renewcommand{\Re}{{\rm Re\ }}          
\renewcommand{\Im}{{\rm Im\ }}          
\newcommand{\bra}[1]{\left\langle#1\right|}  
\newcommand{\ket}[1]{\left|#1\right\rangle}  

\renewcommand{\vec}[1]{\boldsymbol{#1}} 

\newcommand\blfootnote[1]{%
 \begingroup
 \renewcommand\thefootnote{}\footnote{#1}%
 \addtocounter{footnote}{-1}%
 \endgroup
}                                      

\newcommand{\appsection}[2]{\setcounter{equation}{0}\setcounter{subsection}{0}\setcounter{table}{0}\setcounter{figure}{0}
\section*{Appendix #1. #2}
\renewcommand{\theequation}{#1.\arabic{equation}}
              \renewcommand{\thesection}{#1}\renewcommand{\thetable}{#1\arabic{table}}\renewcommand{\thefigure}{#1.\arabic{figure}} }

\begin{document}

\begin{titlepage}

\vskip 1.5 cm
\begin{center}
{\Large \bf Generalised fractional Rabi problem}
\end{center}

\vskip 2.0 cm
\centerline{{\bf Alexander L\'opez}$^{a}$, {\bf S\'ebastien Fumeron}$^{b}$, {\bf Malte Henkel}$^{b,c}$, {\bf Trifce Sandev}$^{d,e,f}$ and {\bf Esther D. Guti\'errez}$^{a}$}
\vskip 0.5 cm
\begin{center}
$^a$Escuela Superior Polit\'ecnica del Litoral, ESPOL,\\ Departamento de F\'isica, Facultad de Ciencias Naturales y Matem\'aticas,\\ Campus Gustavo Galindo
 Km. 30.5 V\'ia Perimetral, P. O. Box 09-01-5863, Guayaquil, Ecuador \\
 $^b$Laboratoire de Physique et Chimie Th\'eoriques (CNRS UMR 7019),\\  Universit\'e de Lorraine Nancy,
B.P. 70239, F -- 54506 Vand{\oe}uvre l\`es Nancy Cedex, France\\
$^c$Centro de F\'{i}sica Te\'{o}rica e Computacional, Universidade de Lisboa, \\Campo Grande, P--1749-016 Lisboa, Portugal\\
$^d$Research Center for Computer Science and Information Technologies, Macedonian Academy of Sciences and Arts, Bul. Krste Misirkov 2, 1000 Skopje, Macedonia\\
$^e$Institute of Physics, Faculty of Natural Sciences and Mathematics,
Ss. Cyril and Methodius University, Arhimedova 3, 1000 Skopje, Macedonia\\
$^f$Department of Physics, Korea University, Seoul 02841, Republic of Korea
\end{center}
\blfootnote{Corresponding author: Alexander Lopez $\text{alexlop@espol.edu.ec}$}

\begin{abstract}
Fractional quantum dynamics provides a natural framework to capture nonlocal temporal behavior and memory effects in quantum systems. In this work, we analyze the physical consequences of fractional-order quantum evolution using a Green’s function formulation based on the Caputo fractional derivative. Explicit iterative expressions for the evolved state are derived and applied to an extended two-level Rabi model, a paradigmatic setting for coherent quantum control. We find that even in the absence of external driving, the static Hamiltonian term induces non-trivial spin dynamics with damping features directly linked to the fractional temporal nonlocality. When a periodically varying driving field is introduced, the competition between energy injection and memory effects gives rise to a richer dynamical behavior, manifest in the evolution of spin polarization, autocorrelation function, and fidelity. Unlike the standard Rabi oscillations characterized by a fixed frequency, the fractional regime introduces controllable damping and dephasing governed by the degree of fractionality. These distinctive signatures could be observable through the Loschmidt echo and autocorrelation function, and would offer potential routes to probe fractional quantum dynamics experimentally. Our findings open pathways toward exploring memory-induced dynamical phenomena in other systems effectively described by a two-level approximation, such as graphene-like materials and topological SSH chains, where non-integer order evolution may reveal novel topological or relaxation effects. \\[0.9cm]~
\centerline{\large \today}
\end{abstract}

\vfill

\end{titlepage}

\setcounter{footnote}{0}

\section{Introduction}

Studies of two-level quantum systems subject to external, time-dependent fields occupy a foundational position in quantum physics, with the Rabi problem providing a classical example of resonant driving and coherent oscillations \cite{Rabi1937}. Moreover, it is one of the few driven systems which admit an exact analytical solution \cite{Shirley1965,Grifoni1998,Chu2004}. Traditionally, the dynamics of these systems are captured by the time-dependent Schrödinger equation with a first-order time derivative and a Hamiltonian that may explicitly depend on time. More precisely, when the Hamiltonian is periodic in time, $H(t)=H(t+T)$, where $T$ is the time period of the periodic interaction, Floquet theory asserts that it is possible to find an equivalent time-independent dynamical evolution in terms of the so called Floquet states and a periodic spectrum of quasienergies. 
Recently, a growing body of research has explored the extension of quantum dynamics to fractional order time derivatives, motivated by the need to describe non-Markovian and memory-dependent phenomena across a variety of physical contexts \cite{Laskin2000original,Laskin2000,Laskin2000chaos,Laskin2002,Naber2004,Tarasov2008,Iomin2009,Iomin2011,Iomin2019,Achar2013,Laskin2017}, including non-Markovian dynamics in open quantum systems~\cite{Tarasov2017,Tarasov2021,Wei2024,Zu2025} and optics \cite{Longhi2015}. 

Scientific interest in fractional calculus has continually grown, with many foundational results presented  in classic texts  \cite{book1,book2,book3,book4,book5,book6}. Previous works have used the fractional calculus approach to analyze spontaneous emission in a two-level atom \cite{Wu2010} and the propagation dynamics of a light beam in a fractional Schr\"odinger equation \cite{Zhang2015}. Moreover, the potential condensed matter realization of fractional dynamics was put forward in reference
\cite{Stickler2013}, within the context of a one-dimensional L\'evy crystal, while the authors of reference \cite{Gabrick2023} have recently discussed the  fractional Schrödinger equation with time dependent potentials. Within this context, the fractional extension of the Floquet theory
\cite{Iomin2023},  which exploits the principle known as subordination, recasting the fractional dynamics in terms of solutions of a related integer-order differential equation.
These studies naturally lead to the so-called fractional Rabi problem, which generalizes the coherent dynamics of a two-level system by replacing the standard time derivative with its fractional analog, typically defined in the sense of Caputo because of its compatibility with physically meaningful initial conditions. The recent experimental demonstration of the fractional Schr\"{o}dinger equation in the temporal domain by Liu and collaborators \cite{Liu2023}
for femtosecond laser pulses, shows that the transmission of input pulses carrying a fractional phase exhibits a “fractional-phase protection” effect through a regular (non-fractional) material, which offers the potential for designing optical signal-processing schemes. 

Theoretically, the fractional Rabi problem provides a rich playground for exploring the interplay between memory, non-locality, and quantum coherence, enabling the study of non-Markovian open quantum systems and anomalous dynamics that cannot be captured by ordinary Schrödinger evolution. Yet, it also introduces several theoretical challenges as the fractional dynamics may be non-unitary, which undermines the traditional probabilistic interpretation of the wave-function and raises questions concerning conservation laws and time-reversal symmetry  \cite{Naber2004,Cius2022}. One approach to the dissipative Rabi oscillations could be by employing the time fractional Lindblad equation without dissipative terms, such that memory-induced weak dissipation in the system occurs~\cite{ang}. Besides, adapting time-dependent perturbation theory to fractional equations is non-trivial, especially for quantum systems described by time-dependent Hamiltonians; new schemes must be constructed that respect the intricacies of fractional calculus the formulation of an appropriate fractional perturbation theory \cite{Jin2021}. 

Beyond its foundational interest, the fractional Rabi problem and its outcomes for quantum dynamics may find applications in several emerging areas. The first class of applications connected to it concerns strongly correlated electron systems and topological matter. Rabi oscillations have indeed been involved in Chern insulators \cite{Zhao2023,Binanti2024} and may help investigate dissipative or memory-effect transport in these media (as a sidenote, the Hamiltonian describing the Rabi oscillation in the presence of time-periodic magnetic field also shares analogous features to that of a 2D Chern insulator \cite{Park2018}, which emphasizes the relevance of our model for topological matter). The physics of topological superconductors is also concerned, as they are natural hosts for fractional excitations and fractional dynamics may help probing hybridization of collective modes, and the unconventional dynamics of Majorana zero modes and anyons \cite{Vaezi2013,Wang2015}. Finally, experiments with ultracold atoms routinely implement the standard Rabi model to study coherent population transfer, entanglement generation, and quantum simulation. Disorder, field inhomogeneities, and interaction with tailored environments introduce non-Markovian effects \cite{Guimond2016} that could be captured by fractional quantum models.

Recent works explore the fractional quantum dynamics of a two-level system in presence of an explicitly time dependent dynamical generator \cite{Lu2016,Lu2017,Lu2018,Zu2021}. In this work we explore a general fractional order time dependent Schr\"odinger equation, generated by explicitly time-dependent Hamiltonians and derive the associated perturbation theory scheme using the Caputo definition of the time-dependent dynamical generator giving the explicit form of the fractional version of the Dyson's series \cite{Tarasov2017}. Yet, since the resulting expansion includes the static Hamiltonian, even the leading order result is not expected to properly capture the emergent dynamical features of the system. For the standard Schr\"odinger equation, this is solved by using the representation picture, which encodes the static Hamiltonian in a phase and allows the dynamics to be dictated by the transformed interaction. This in turn relies on the Leibniz property of the integer order derivative. Since the fractional derivative lacks this property, we introduce the fractional Green's function expansion that only depends on the perturbation potential, and the Green's function associated to the static system. This is one of the main results of this paper. Then, we apply these results to obtain the leading order dynamical response of the fractional Rabi problem for a two-level system and obtain expressions for the time evolution of spin polarization, autocorrelation function and Loschmidt echo. From these quantities we infer the salient physical consequences of periodical driven systems with memory and discuss their potential relevance to gain further insight on the role of memory effects in periodically driven systems. This provides the second major result of this paper as we can use these two quantities to  distinguish  between the early- and long-term dynamical features, where the fidelity at longer times shows a time evolution almost independent from the memory effect, while the autocorrelation function shows a complementary behavior.  The paper is organized as follows: In section 2 we present the perturbative Dyson series and Green's function expansion for the fractional time dependent Schr\"odinger equation. In section 3 we briefly summarize the results for the  dynamics of a two-level system and in section 4 we apply the fractional Green's function approach to solve the memory-dependent fractional Rabi problem. In section 5 we discuss the leading order results for the dynamical evolution of the spin polarization, autocorrelation function and fidelity or Loschmidt echo. We summarize our results in section 6 and give concluding remarks. Some additional calculations are presented in the two appendices.

\section{Perturbative approach to the time-dependent fractional Schr\"odinger equation}

The fractional time Schr\"odinger equation (FTSE) formulated using the Caputo derivative, is defined as 
\begin{equation}\label{ftsq1}
\II\hbar^\alpha D^\alpha_t \Psi(t)=H(t)\Psi(t),  
\end{equation}
where $\hbar=h/2\pi$ is the reduced Planck's constant, $H(t)$ is the explicitly time-dependent Hamiltonian or dynamical generator and $\Psi(t)$ 
is to be interpreted as the wave function within the fractional time realm. In addition, the fractional-order Caputo derivative is defined as~\cite{book3} 
\begin{equation}
D^\alpha_t f(t)=\frac{1}{\Gamma(1-\alpha)}\int_{0}^t\frac{\D t'}{(t-t')^\alpha}\frac{\D f(t')}{\D t'}, \quad 0<\alpha<1.    
\end{equation}
We should mention an alternative approach where the fractional dynamics is given by
\begin{equation}\label{ftsq2}
(\II\hbar)^\alpha D^\alpha_t \Psi(t)=H(t)\Psi(t), 
\end{equation}
which has been extensively analyzed in references \cite{Naber2004,Iomin2009,Gabrick2023}.


\noindent Before addressing the time-dependent case, let us recall the restricted situation of a static Hamiltonian $H(t)=H_s$, one needs to solve the eigenvalue problem
\begin{equation}\label{static}
\II\hbar^\alpha D^\alpha_t \psi(t)=H_s\psi(t).  
\end{equation}
We can write the quantum state as a linear combination of the eigenstates of the static Hamiltonian, $H_s\ket{\phi_\sigma}=\lambda_\sigma\ket{\phi_\sigma}$, in the form \begin{equation}
\ket{\psi(t)}=\sum_\sigma c_\sigma(t)\ket{\phi_\sigma},    
\end{equation}
which upon using the orthogonality relation $\langle\phi_{\sigma'}\ket{\phi_\sigma}=\delta_{\sigma'\sigma}$ leads to the dynamical evolution equation for the time-dependent expansion coefficients
\begin{equation}
\II\hbar^\alpha D^\alpha_t c_\sigma(t)=\lambda_\sigma c_\sigma(t)  
\end{equation}
Thus, we can write the solution as  
\begin{equation}\label{static-sol}
c_\sigma(t)=c_\sigma E_\alpha\left(-\frac{\II\lambda_\sigma t^\alpha}{\hbar^\alpha}\right),
\end{equation}
where $c_\sigma=c_\sigma(0)$ its the initial value for the expansion coefficient and the one-parameter Mittag-Leffler function is defined as (see, for example, Ref.~\cite{book3}),
\begin{equation}\label{ML-static}
E_\alpha(z)=\sum_{n=0}^\infty\frac{z^n}{\Gamma(\alpha n+1)}.  
\end{equation}
In general, for the explicitly time-dependent fractional Schr\"odinger equation, exact analytical solutions would be scarce to find. Thus, we write now a self-consistent scheme that could afford approximate dynamical solutions to the FTSE in eq. (\ref{ftsq1}). For this purpose, we write it as
\begin{equation}
\label{eq_caputo}
D^\alpha_t \ket{\Psi_\alpha(t)}=\lambda(t)\ket{\Psi_\alpha(t)},    
\end{equation}
with $\lambda(t)=-\II\hbar^{-\alpha}H(t)$. 
Using the definition of the Caputo derivative at $t=t_1$, we get 
\begin{equation}
\frac{1}{\Gamma(1-\alpha)}\int_0^{t_1}\frac{\D t_2}{(t_1-t_2)^\alpha}\frac{\D\Psi(t_2)}{\D t_2}=\lambda(t_1)\Psi(t_1).
\end{equation}
Now we multiply both sides by the kernel $(t-t_1)^{\alpha-1}$ and integrate in the interval $0\leq t_1\leq t$, to get
\begin{equation}
\frac{1}{\Gamma(1-\alpha)}\int_0^{t}\frac{\D t_1}{(t-t_1)^{1-\alpha}}\int_0^{t_1}\frac{\D t_2}{(t_1-t_2)^\alpha}\frac{\D \Psi(t_2)}{\D t_2}=\int_0^{t}\frac{\D t_1}{(t-t_1)^{1-\alpha}}\lambda(t_1)\Psi(t_1).
\end{equation}
and make use of Fubini's theorem, such that we can rewrite the left hand side of this expression as
\begin{equation}\label{13}
\frac{1}{\Gamma(1-\alpha)}\int_0^{t}\D t_2\frac{\D\Psi(t_2)}{\D t_2}\int_{t_2}^{t_1}\frac{\D t_1}{{(t-t_1)^{1-\alpha}}(t_1-t_2)^\alpha}=\int_0^{t}\frac{\D t_1}{(t-t_1)^{1-\alpha}}\lambda(t_1)\Psi(t_1).
\end{equation}
Now we invoke the result (see Appendix A for details of the derivation)
\begin{equation}
I_{\alpha\beta}(x,z)=\int_z^x\frac{\D y}{(x-y)^\alpha(y-z)^\beta}=(x-z)^{1-(\alpha+\beta)}\frac{\Gamma(1-\alpha)\Gamma(1-\beta)}{\Gamma(2-(\alpha+\beta))}, 
\end{equation}
and we recognize that the second integral on the left-hand side of eq. (\ref{13}) corresponds to $I_{\alpha,1-\alpha}(t_1,t_2)=\Gamma(\alpha)\Gamma(1-\alpha)$. Then, we get
\begin{equation}
\Gamma(\alpha)\int_0^{t}\D t_2\frac{\D \Psi(t_2)}{\D t_2}=\int_0^{t}\frac{\D t_1}{(t-t_1)^{1-\alpha}}\lambda(t_1)\Psi(t_1),
\end{equation}
from which we obtain the solution as
\begin{equation}
\Psi(t)=\Psi(0)+\frac{1}{\Gamma(\alpha)}\int_0^t\frac{\D t_1\lambda(t_1)\Psi(t_1)}{(t-t_1)^{1-\alpha}}.
\end{equation}
In terms of the Hamiltonian, the dynamics read as
\begin{equation}
\Psi(t)=\Psi(0)+\frac{1}{\II\gamma_\alpha}\int_0^t\frac{\D t_1}{(t-t_1)^{1-\alpha}} H(t_1)\Psi(t_1)=\Psi(0)+\frac{1}{\II\hbar^\alpha}I_{t}^{\alpha}H(t)\Psi(t),
\end{equation}
where $\gamma_\alpha=\hbar^\alpha\Gamma(\alpha)$ and
\begin{align}
    I_{t}^{\alpha}=\frac{1}{\Gamma(\alpha)}\int_{0}^{t}\frac{f(t')}{(t-t')^{1-\alpha}} \D t', \quad 0<\alpha<1, \quad t>0,
\end{align}
is the Riemann-Liouville fractional integral~\cite{book3}. This equation can be obtained if we apply the Riemann-Liouville fractional integral on both sides of eq.~(\ref{eq_caputo}), and then use the relation $I_t^{\alpha}D_t^{\alpha}f(t)=f(t)-f(0)$. 
This equation can be iterated to get,
\begin{eqnarray}
\Psi(t)&=&\Psi(0)+\frac{1}{\II\gamma_\alpha}\int_0^t\frac{\D t_1}{(t-t_1)^{1-\alpha}} H(t_1)\Psi(0)+\nonumber\\
&&\left(\frac{1}{\II\gamma_\alpha}\right)^2\int_0^t\frac{\D t_1}{(t-t_1)^{1-\alpha}} H(t_1)\int_0^{t_1}\frac{\D t_2}{(t_1-t_2)^{1-\alpha}}H(t_2)\Psi(t_2).\nonumber\\
\end{eqnarray}
Formally speaking, we can write this solution as
\begin{equation}
\Psi(t)=\left[\mathbf{1}+\frac{1}{\II\gamma_\alpha}\int_0^t\frac{\D t_1}{(t-t_1)^{1-\alpha}} H(t_1)+\left(\frac{1}{\II\gamma_\alpha}\right)^2\int_0^t\frac{\D t_1}{(t-t_1)^{1-\alpha}} H(t_1)\int_0^{t_1}\frac{\D t_2}{(t_1-t_2)^{1-\alpha}}H(t_2)\cdots\right]\Psi(0),
\end{equation}
which we can further rewrite in the suggestive form of a time-ordered ``exponential''\cite{Tarasov2017}:
\begin{equation}
\ket{\Psi_\alpha(t)}=\mathcal{T}\exp\left(\frac{1}{\II\gamma_\alpha}\int_0^t\frac{\D t_1}{(t-t_1)^{1-\alpha}} H(t_1)\right)\ket{\Psi(0)} =:  U_\alpha(t)\Psi(0),
\end{equation}
where we have introduced the effective evolution operator
\begin{equation}
U_\alpha(t)=\mathcal{T}\exp\left(\frac{1}{\II\gamma_\alpha}\int_0^t\frac{\D t_1}{(t-t_1)^{1-\alpha}} H(t_1)\right).
\end{equation}
As expected, in the limit $\alpha\rightarrow 1$, we recover the standard result for the evolution operator  
\begin{equation}
U(t)=\mathcal{T}\exp\left(\frac{1}{\II\hbar}\int_0^t \!\D t_1\: H(t_1)\right).
\end{equation}
This result is exact in the sense that no approximations have been made so far. 
However,  this form is not directly   suitable for a perturbative treatment as by neglecting higher order terms we could miss important features of the actual dynamical evolution of the system. This stems  from the fact that the static Hamiltonian typically has a larger influence on the dynamics than the perturbation potential, at least within the regime of validity of perturbation theory. 

\noindent For standard quantum mechanics, the Leibniz rule for products of function allows the formulation in the interaction picture. 
Then, approximate solutions can be readily found. In the fractional calculus formulation, that approach would not work since the Leibniz rule becomes cumbersome due to its combinatorial nature 
(see for instance references \cite{book1,book2,book3,book4,book5,book6}). Thus, we rewrite the solution following a different strategy.  

\noindent In general, the Hamiltonian could be realised to consist of a static part plus a time-dependent contribution in the form
\begin{equation}
H(t)=H_s+V(t),
\end{equation}
where the static contribution $H_s$ to the total Hamiltonian $H(t)$ admits a set of solutions to the static evolution 
\begin{equation}
\II \hbar^\alpha D_t^\alpha\ket{\psi_\alpha(t)}=H_s\ket{\psi_\alpha(t)},
\end{equation}
as it was described in eq.~(\ref{static-sol}). 
Thus, if we rewrite the dynamical evolution as
\begin{equation}\label{driven2}
(\II\hbar^\alpha D^\alpha_t-H_s) \ket{\Psi_\alpha(t)}=V(t)\ket{\Psi_\alpha(t)},   
\end{equation}
and introducing the fractional Green's function $G_\alpha(t,t')$, which is defined as a solution of the dynamical equation
\begin{equation}
(\II\hbar^\alpha D^\alpha_t-H_s)G_\alpha(t,t')=\delta(t-t'),    
\end{equation}
with $\delta(t-t')$ being the Dirac delta function, which in turn is defined in the standard form as
\begin{equation}
\int\delta(t-t')f(t')\D t'=f(t).
\end{equation}
Therefore, we can formally write the solution to the dynamical evolution (\ref{driven2}) in the self-consistent form
\begin{equation}\label{driven3}
\ket{\Psi_\alpha(t)}=\ket{\psi_\alpha(t)}+\int \!\D t'\: G_\alpha(t,t')V(t')\ket{\Psi_\alpha(t')},    
\end{equation}
with $\ket{\psi_\alpha(t)}$ being the solution generated by the static Hamiltonian $H_s$, in eq.~(\ref{static-sol}).
A comment about the dimensionality of the fractional Green's function is in order. For the standard Schr\"odinger equation $[G(t)]=[\hbar^{-1}]$ (inverse of action). 
Within the fractional regime, one has that its dimensions are 
$[G_\alpha(t)]=[1/(tE^\alpha)]\equiv [t^{\alpha-1}\hbar^{-\alpha}]$, such that $\lim_{\alpha\rightarrow1}[G_\alpha(t)]=[\hbar^{-1}]$.

\noindent The general solution of the fractional dynamical problem is obtained as the series
\begin{equation}\label{expansion}
\ket{\Psi_\alpha(t)}=\ket{\psi_\alpha(t)}+\int \!\D t'\:G_\alpha(t,t')V(t')\ket{\psi_\alpha(t')}+\int \!\D t'\:G_\alpha(t,t')V(t')\ket{\psi_\alpha(t')}\int \!\D t''\: G_\alpha(t',t'')V(t'')\ket{\psi_\alpha(t'')}\cdots    
\end{equation}
For suitable parameter regimes, i.e., when the perturbation strength has an energy scale much smaller than those associated with the static Hamiltonian, the leading-order solution to the fractional time-dependent Schr\"odinger equation  is given by 
\begin{equation}\label{leading}
\ket{\Psi_\alpha(t)}\approx\ket{\psi_\alpha(t)}+\int G_\alpha(t,t')V(t')\ket{\psi_\alpha(t')}\D t'.    
\end{equation}
Using these general results, in the following section we explore the dynamical features of a driven two-level system, described by an extension of the fractional time Schr\"odinger equation.

\section{Dynamical quantities of interest of a two-level system} 
Let us apply the previous results to the paradigmatic case of a two-level system (TLS). In this case, the solution to the Schr\"odinger equation
$\II \hbar D_t\ket{\Psi(t)}=H(t)\ket{\Psi(t)}$, can be written as
\begin{equation}
\ket{\Psi(t)}=\left(
\begin{array}{cc}
     c_+(t) \\
     c_-(t) 
\end{array}\right).
\end{equation}
From this solution, we can calculate the spin polarization evolution, which is defined as \begin{equation}\label{spinpol}
\langle\sigma_j(t)\rangle=\bra{\Psi(t)}\sigma_j\ket{\Psi(t)}\qquad j=x,y,z.    
\end{equation}
Performing the explicit calculations, we get the results
\begin{eqnarray}
\langle \sigma_x(t)\rangle&=&2\Re[c^{*}_+(t)c_-(t)],\\
\langle \sigma_y(t)\rangle&=&2\Im[c^{*}_+(t)c_-(t)],\\
\langle \sigma_z(t)\rangle&=&|c_+(t)|^2-|c_-(t)|^2.
\end{eqnarray}
Another physical quantity of interest is the autocorrelation function, which is defined as
\begin{equation}
A(t)=|\langle\Psi(0)\ket{\Psi(t)}|^2.   
\end{equation}
Therefore, it measures the probability that the evolved system returns to its initial configuration up to a phase. The times at which $C(t_R)=1$ are called revival times, while the times when it achieves a local maximum smaller than 1, are called partial revival times. In addition, the imaginary part of the Fourier transform of the autocorrelation function is proportional to the density of states, which in turn measures the probability of occupation of an energy eigenstate\cite{Robinett2004,Krueckl2009,Romera2009, Romera2011,Lopez2015}. 
Its explicit calculation gives
\begin{equation}
A(t)=|c^{*}_+(0)c_+(t)+c^{*}_-(0)c_-(t)|^2.    
\end{equation}
As another measure of the memory effects associated to the fractional evolution dynamics, we can use fidelity, which in the context of chaos theory is also termed a Loschmidt echo\cite{Goussev2012}, which is defined as 
\begin{equation}
F(t)=|\langle{\Psi_p}(t)\ket{\Psi(t)}|^2=|p^{*}_+(t)c_+(t)+p^{*}_-(t)c_-(t)|^2,
\end{equation}
where $\ket{\Psi_p(t)}=p_+\ket{+}+p_-\ket{-}$.
From its definition, fidelity measures  how a perturbation affects the dynamics of an initially prepared  state $\ket{\Psi(0)}$ that evolves forward in time according to a known dynamical generator, given the state $\ket{\Psi(t)}$, and then evolves backward in time with the dynamical generator that includes the perturbation, leading to the state $\ket{\Psi_p(t)}$. It is important to notice that when the initial state is an eigenstate of the unperturbed Hamiltonian, the fidelity and autocorrelation function coincide.   
These quantities would be the observables of interest that encode the information of the memory effects and could render a possible experimental realization of our results\cite{Martinez2016,Jurcevic2017,Braumueller2022,Karch2025}. 
\section{Application to generalised Rabi problem}
A paradigmatic example of TLS is the standard Rabi problem, for which we have the following Hamiltonian 
\begin{equation}\label{Rabi}
H_R(t)=\Delta\sigma_z+\xi(\sigma_x\cos{\Omega t}+\sigma_y\sin{\Omega t})\equiv\vec{\sigma}\cdot\vec{h}_{R}(t),    
\end{equation}
where $\vec{h}_R(t)=(\Delta,\xi\cos\Omega t,\xi\sin\Omega t)$, which describes the dynamical evolution of a spin $1/2$ particle coupled to a magnetic field that has a constant component along the $z$ axis, and an oscillatory field in the $x-y$ plane, of frequency $\Omega$. The Pauli spin matrices have been arranged in a vector $\vec{\sigma}=(\sigma_x,\sigma_y,\sigma_z)$, while $\Delta$ gives the static energy splitting, while $\xi$ measures the effective coupling strength to the oscillatory component of the periodically varying magnetic field. As is well known, the Schr\"odinger equation for the standard time-dependent Rabi problem 
\begin{equation}\label{Rabi-integer}
\II\hbar D_t\ket{\Psi(t)}=H_R(t)\ket{\Psi(t)},    
\end{equation}
is exactly solvable by means of an appropriate time-dependent unitary transformation. Thus, it constitutes one of the few dynamical problems admitting an exact analytical closed form (see the appendix for an explicit derivation of the solution), which reads 
\begin{equation}
\ket{\Psi(t)}=e^{-\II\Omega t\sigma_z/2}e^{-\II H_F t/\hbar}\ket{\Psi(0)},    
\end{equation}
where we have defined
\begin{equation}
H_F=(\Delta-\hbar\Omega/2)\sigma_z+\xi\sigma_x.
\end{equation}

We now analyze the dynamics of an extension of this Rabi Hamiltonian within the fractional quantum-mechanical formulation. Among the potential extensions of the Rabi Hamiltonian to the fractional regime, we consider the following Hamiltonian with dimensionally corrected parameters
\begin{equation}
H_{R,\alpha}(t)=\Delta^\alpha\sigma_z+\xi^\alpha(\sigma_x\cos{\Omega t}+\sigma_y\sin{\Omega t})\equiv\vec{\sigma}\cdot\vec{h}_{R,\alpha}(t), 
\end{equation}
As expected, this leads to the standard Rabi Hamiltonian as $\alpha\rightarrow 1$. Let us analyze the main emergent dynamical features, first focusing on the role of the memory effects on the spin polarization for the static contribution to the modified Rabi problem.
In this case, we identify $H_s=\Delta^\alpha\sigma_z$ which is diagonal, and thus the corresponding solutions $\psi(t)$ are given in terms of the Mittag-Leffler function, which read explicitly as:
\begin{equation}\label{rabi1}
\ket{\psi_\alpha(t)}=E_\alpha\left(-\II\omega^\alpha t^\alpha\sigma_z\right)\ket{\psi(0)},   
\end{equation}
where $\hbar\omega=\Delta$.
To write the initial state $\psi(0)$, a convenient parametrisation of the expansion coefficients can be implemented by means of the polar angle $\theta$, whose variation in the Bloch sphere is the domain $0\leq\theta<\pi$, corresponding to a continuous transition from the eigenstate $\ket{+}$, located at the north pole of the sphere, to $\ket{-}$, which lies at the south pole of the Bloch sphere. Thus, we write the initial state as   
\begin{equation}\label{initial-theta}
\ket{\psi(0)}=\cos\theta/2\ket{+}+\sin\theta/2\ket{-}.   \end{equation}\label{initial}
\begin{figure}
\begin{center}    \includegraphics[height=6cm]
{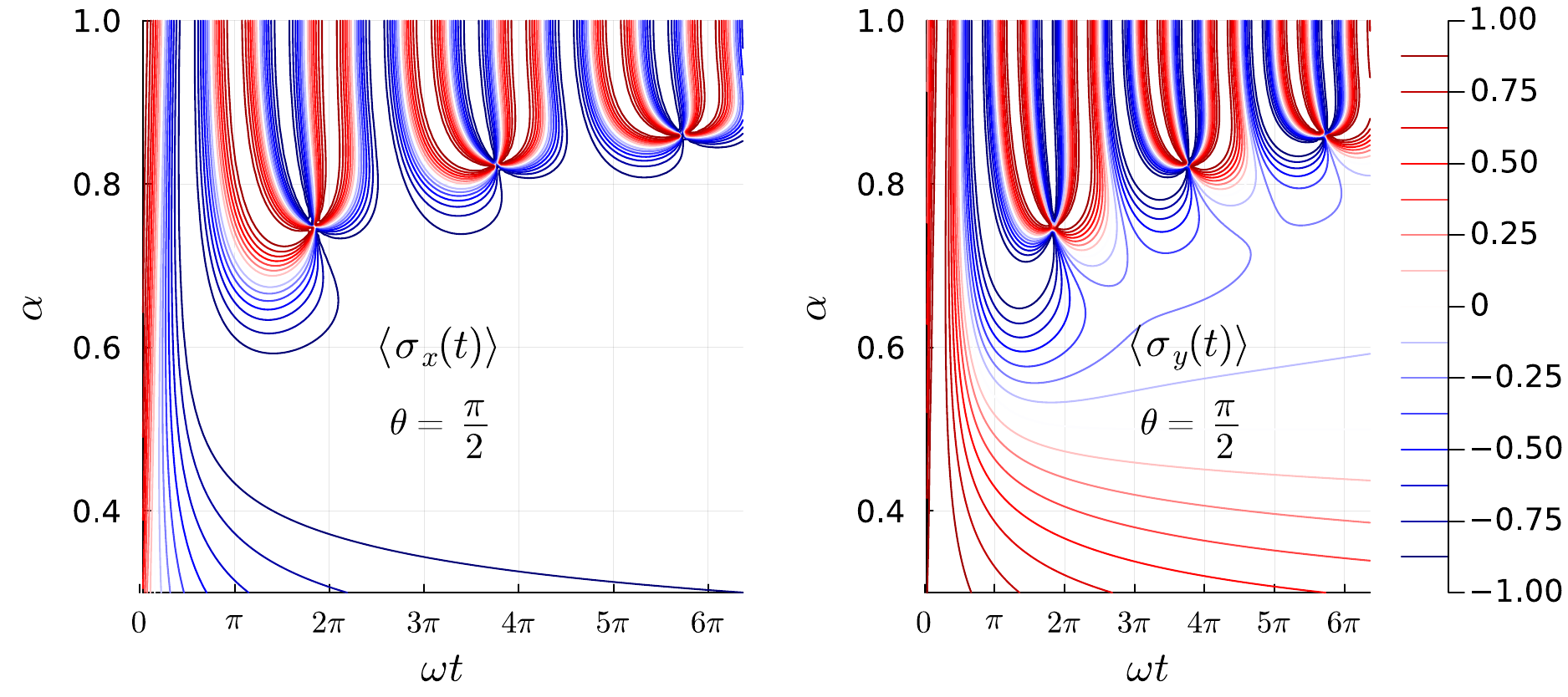}
\end{center}
\caption{ Contour plots for the time evolution of spin polarization components $\langle \sigma_x(t) \rangle$, and $\langle \sigma_y(t) \rangle$, under a static fractional Hamiltonian. As $\alpha$ decreases from $1$ to lower values, the system exhibits slower relaxation and enhanced memory effects, with separated open and closed contours. These results illustrate how fractional dynamics alter the coherence and population evolution in comparison to the standard ($\alpha = 1$) unitary case.}
\end{figure}
Using the solution given in eq. (\ref{rabi1}), we assess the time evolution of the spin polarization  components, which are defined through the expressions given in eqs. \ref{spinpol}.
Performing the explicit calculations, one gets
\begin{eqnarray}
\langle\sigma_x(t)\rangle&=&\frac{\Re[E^{2}_\alpha(\II\omega^\alpha t^\alpha)]\sin\theta}{|E_\alpha(\II\omega^\alpha t^\alpha)|^2}\label{sigmax},\\
\langle\sigma_y(t)\rangle&=&\frac{\Im[E^{2}_\alpha(\II\omega^\alpha t^\alpha)]\sin\theta}{|E_\alpha(\II\omega^\alpha t^\alpha)|^2}\label{sigmay},\\
\langle\sigma_z(t)\rangle&=&\cos\theta\label{sigmaz}.
\end{eqnarray}
As expected, if the initial state is an eigenstate of $\sigma_z$, meaning that one of the coefficients is zero, say $\theta=0$, then only eq. (\ref{sigmaz}) gives a non-vanishing result. On the other hand, when the two coefficients have the same magnitude, i.e., the initial prepared state is an eigenstate of either $\sigma_x$ or $\sigma_y$, eq. (\ref{sigmaz}) exactly vanishes and eqs. (\ref{sigmax}) and \ref{sigmay}) achieve their maxima at $t=0$, and oscillate in time. These results are shown in (\figurename{1}), where we have fixed $\theta=\pi/2$, that is, along the equatorial line of the Bloch sphere, to assess the time evolution of $\langle\sigma_x(t)\rangle$ and $\langle\sigma_y(t)\rangle$. 
From the panels of \figurename{1}, one observes the strong memory effects associated with the values of the fractional-order parameter $\alpha\leq0.5$, and partial spin oscillations in the $x-y$ plane are achieved around $0.6\leq\alpha\leq0.9$. In fact, unless $\alpha=1$ the time oscillations of the $x-y$ spin components will decay in amplitude for a long enough time of observation. Interestingly, the fractional dynamics distinguishes well separated closed and open contours. The open contours show a decaying behavior which is archetypical for the power law asymptotic regime of the Mittag Leffler, while the closed contours show a periodicity that corresponds to half the period of the quantum revivals, as is explicitly seen below when we analyze the autocorrelation function and fidelity.

\noindent Using the result given in eq. (\ref{rabi1}), we now  calculate the leading- order evolution state, which is generated by the full Hamiltonian, i.e., including the perturbation $V(t)$. Thus, we need to determine the Green's function $G_\alpha(t,t')\equiv G(t-t')$, which would be found by solving the fractional evolution equation
\begin{equation}
(\II\hbar^\alpha D_t^\alpha-\Delta^\alpha\sigma_z)G(t-t')=\delta(t-t').    
\end{equation}
Since the static Hamiltonian is diagonal, the associated Green's function is also diagonal, and we can rewrite this in the convenient form
\begin{equation}
(\II\hbar^\alpha D_t^\alpha-\sigma\Delta^\alpha)G_\sigma(t-t')=\delta(t-t'), 
\end{equation}
where $\sigma=\pm1$. 
To solve this equation, we use the result
\begin{equation}\label{transform}
\int_{t'}^t K(t-t'')f(t''-t')\D t''=\int_{0}^{\tau}K(\tau-\tau_1)f(\tau_1)\D\tau_1,    
\end{equation}
where $\tau=t-t'$.
We now write explicitly the evolution equation for the Green's function, by using the definition of the Caputo derivative
\begin{equation}
\frac{\II\hbar^\alpha}{\Gamma(1-\alpha)}\int_0^\tau\frac{\D\tau_1}{(\tau- \tau_1)^\alpha}\frac{\D G_\sigma(\tau_1)}{\D\tau_1}-\sigma\Delta^\alpha G_\sigma(\tau)=\delta(\tau), 
\end{equation}
Using the definition of the fractional integration operator
\begin{equation}
I_\alpha[f(\tau)]=\frac{1}{\Gamma(\alpha)}\int_0^\tau\frac{\D \tau'}{(\tau-\tau')^{1-\alpha}}f(\tau')    
\end{equation}
which is the inverse of the fractional derivative $I_\alpha[D^\alpha_\tau f(\tau)]=f(\tau)$, we get
\begin{equation}
\II\hbar^\alpha G_\sigma(\tau)-\frac{\sigma\Delta^\alpha}{\Gamma(\alpha)}
\int_0^\tau \frac{\D \tau_1G_\sigma(\tau_1)}{(\tau-\tau_1)^{1-\alpha}}=\frac{\tau^{\alpha-1}}{\Gamma(\alpha)}.
\end{equation}
Laplace transformation for each term, leads to
\begin{equation}
g_\sigma(s)+\frac{\II\omega^\alpha}{s^\alpha}\sigma g_\sigma(s)=\frac{1}{\II\hbar^\alpha s^\alpha},   \end{equation}
where $g_\sigma(s)=\mathcal{L}[G(\tau)]$  is the Laplace transform for the corresponding component of the Green's function,
defined as
\begin{equation}
 g_\sigma(s)=\int_0^\infty e^{-s\tau}G_\sigma(\tau)\D\tau.   
\end{equation}
This can be rewritten as
\begin{equation}
g_\alpha(s)=\frac{1}{\II\hbar^\alpha}\frac{1}{s^\alpha+\II \sigma\omega^\alpha}.    
\end{equation}
Using the well-known Laplace transform formula
\begin{equation}
\mathcal{L}[\tau^{\beta-1}E_{\alpha\beta}(\gamma\tau^\alpha)]= \frac{s^{\alpha-\beta}}{s^\alpha-\gamma},   
\end{equation}
we get
\begin{equation}
G(t-t')=\frac{E_{\alpha,\alpha}(-\II\omega^\alpha(t-t')^\alpha\sigma_z)}{\II\hbar^\alpha(t-t')^{1-\alpha}},    
\end{equation}
where the two-parameter Mittag-Leffler function is defined as~\cite{book3}
\begin{equation}
E_{\alpha\beta}(z)=\sum_{\ell=0}^\infty\frac{z^\ell}{\Gamma(\ell\alpha+\beta)}.    
\end{equation}
From this result, the dynamical evolution of the system is given to leading order in $\lambda$, as
\begin{equation}
\ket{\Psi_\alpha(t)}=\ket{\psi_\alpha(t)}+\lambda^\alpha\int_0^t \frac{E_{\alpha,\alpha}(-\II\omega^\alpha(t-t')^\alpha\sigma_z)}{\II[\omega(t-t')]^{1-\alpha}}[\sigma_x\cos(\Omega t')+\sigma_y\sin(\Omega t')]\ket{\psi_\alpha(t')} \omega \D t',     
\end{equation}\label{corrected}
where $\ket{\psi(t)}$ is given by eq.~(\ref{rabi1}) and we have set $\lambda=\xi/\Delta$ as the effective dimensionless perturbation strength.
\section{Modulation effects due to periodic driving}
Using the state given in eq. (\ref{corrected}), along with the initial state given in eq. (\ref{initial}), we get the spin polarization evolution to leading order in $\lambda$ given by

\begin{eqnarray}
\langle\sigma_x(t)\rangle&\approx&\sin\theta\frac{\Re[E^{2}_\alpha(\II\omega^\alpha t^\alpha)]}{|E_\alpha(\II\omega^\alpha t^\alpha)|^2}+2\frac{\Re[E_\alpha(-\II\omega^\alpha t^\alpha)P_\alpha(t)]}{|E_\alpha(\II\omega^\alpha t^\alpha)|^2}\label{sigmax2},\\
\langle\sigma_y(t)\rangle&\approx&\sin\theta\frac{\Im[E^{2}_\alpha(\II\omega^\alpha t^\alpha)]}{|E_\alpha(\II\omega^\alpha t^\alpha)|^2}-2\frac{\Im[E_\alpha(-\II\omega^\alpha t^\alpha)P_\alpha(t)]}{|E_\alpha(\II\omega^\alpha t^\alpha)|^2}\label{sigmay2},\\
\langle\sigma_z(t)\rangle&\approx&\cos\theta\label{sigmaz2},
\end{eqnarray}
where
\begin{equation}\label{palpha}
P_\alpha(t)=\lambda^\alpha\int_0^t \frac{E_{\alpha,\alpha}(-\II\omega^\alpha(t-t')^\alpha)}{\II[\omega(t-t')]^{1-\alpha}}E_\alpha(\II\omega^\alpha t'^\alpha)e^{\II\Omega t'}\omega \D t'.    
\end{equation}
\begin{figure}[ht]
\begin{center}    \includegraphics[height=7cm]
{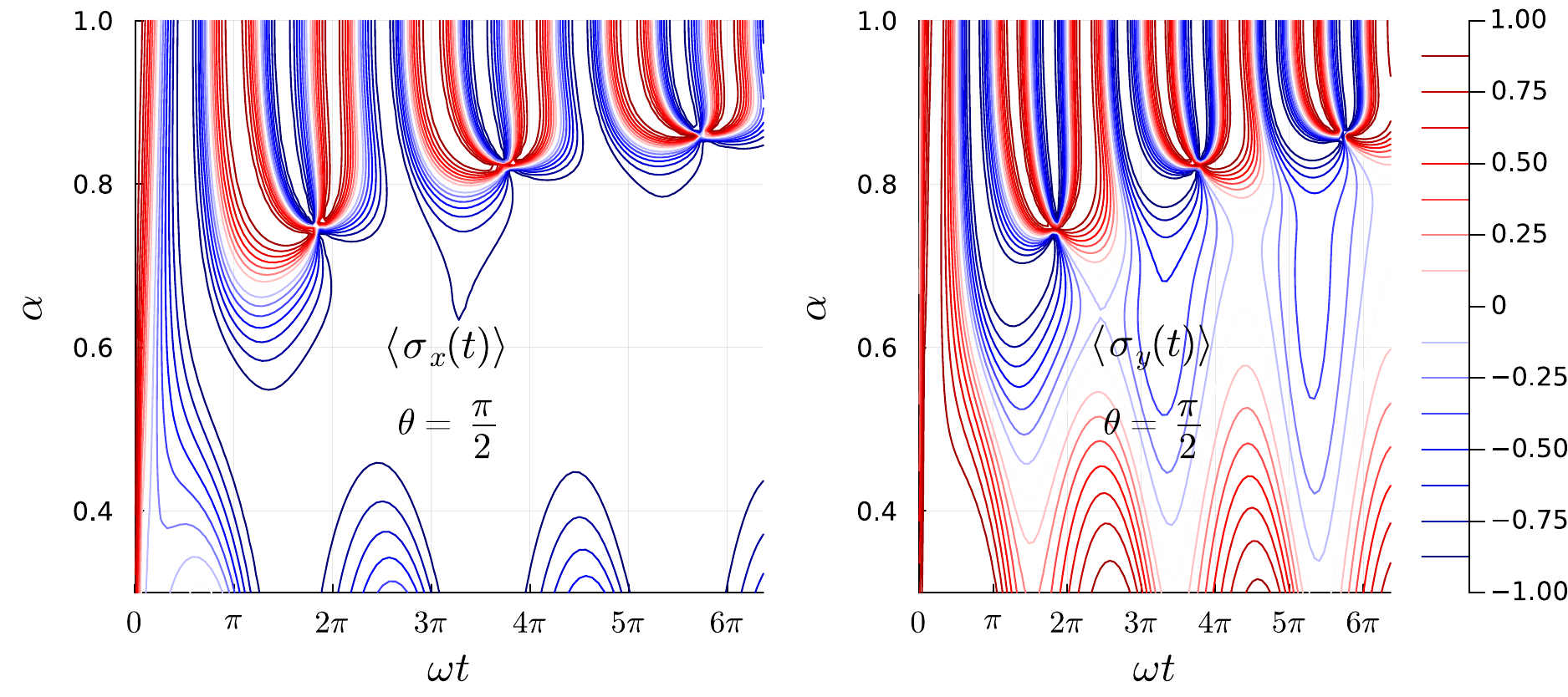}
\end{center}
\caption{Contour plots of the time evolution for the spin polarization components $\langle \sigma_x(t) \rangle$ and $\langle \sigma_y(t) \rangle$ for a two-level system governed by a fractional time evolution with a Mittag-Leffler memory kernel. The fractional order is $\alpha \in [0.2, 1.0]$. The system is subjected to a harmonic external perturbation with frequency $\Omega = \Delta/\hbar$, and the coupling strength is fixed at $\lambda = 0.1$. }
\end{figure}
Therefore, in the leading-order approximation, the $z$ component of the spin polarization is not affected by the driving field, whereas the $x-y$ components get a correction which is independent of the initial-state configuration. 
The apparently singular integral can be solved by expressing the integrand as a product of convergent series, given by
\begin{equation}
P_\alpha(t)=-\II\lambda^\alpha (\omega t)^\alpha e^{\II\Omega t}\sum_{m=0}^\infty\frac{(-\II\Omega t)^m}{m!}\sum_{n=0}^\infty\frac{(\II\omega^\alpha t^\alpha)^n}{\Gamma(\alpha n+\alpha)} {\Gamma(\alpha(n+1)+m)} E_{\alpha,{\alpha(n+1)+m+1}}(-\II\omega^\alpha t^\alpha)    
\end{equation}

The resulting spin dynamics as a function of time and the fractional order parameter is explored graphically in (\figurename{2}) where we have set $\theta=\pi/2$ to compare with the results shown in  (\figurename{1}). The main salient feature of the driving effects is observed in the structure of the  open countours which were present in (\figurename{1}), and now have become closed contours. This effect is a clear signature of the competition between the energy input of the driving and  the dissipative effects associated to the memory properties of the system, which is encoded in the values of $\alpha$. From the evolution of $\langle\sigma_y(t)\rangle$ we infer that the minima and maxima show an alternating pattern. To get further insight on the nature of these minima and maxima, we explore the fidelity and autocorrelation function which are experimental tools that could shed light on the spin polarization effects due to the competition of the memory effects and periodic driving. 

 For the fidelity, we get after lengthy calculations (we refer the reader to Appendix B for a detailed derivation):
 \begin{eqnarray}
F(t)&=&\frac{
 1+(|\tilde{c}_+|^2-|\tilde{c}_-|^2)[\left(|g_+|^2-|g_-|^2\right)\cos\gamma+2\Re(g_+g^{*}_-e^{\II \Omega t})\sin\gamma]}{2}+\nonumber\\
&& -2\Im(\tilde{c}^{*}_+\tilde{c}_-e^{2\II \epsilon t/\hbar})\Im (e^{\II \Omega t}g_+g^{*}_-)+2\Re(\tilde{c}^{*}_+\tilde{c}_-e^{2\II \epsilon t/\hbar})\left[\Re(e^{\II \Omega t}g_+g^{*}_-)\cos\gamma
-\frac{(|g_+|^2-|g_-|^2)}{2}\sin\gamma\right].\nonumber\\
\end{eqnarray}
with $\tan\gamma=2\lambda\omega/(2\omega-\Omega)$, while  $\epsilon=\hbar\sqrt{(2\omega-\Omega)^2+4(\lambda\omega)^2}/2$ is the quasi-energy, which is defined up to $\mod \hbar\Omega$, and can be restricted to the first Floquet Brillouin zone $-\hbar\Omega/2\le\epsilon\le\hbar\Omega/2$. 
For the chosen initial state given in eq. (\ref{initial-theta}), the time evolution of the fidelity reads
\begin{eqnarray}
F(t)&=&\frac{
 1+\cos(\theta-\gamma)[\left(|g_+|^2-|g_-|^2\right)\cos\gamma+2\Re(g_+g^{*}_-e^{\II \Omega t})\sin\gamma]}{2}+\nonumber\\
&& -\sin(\theta-\gamma)\left\{\sin{(2\epsilon t/\hbar})\Im (e^{\II \Omega t}g_+g^{*}_-)-\cos {(2\epsilon t/\hbar)}\left[\Re(e^{\II \Omega t}g_+g^{*}_-)\cos\gamma
-\frac{(|g_+|^2-|g_-|^2)}{2}\sin\gamma\right]\right\}.\nonumber\\
\end{eqnarray}

\begin{figure}
    \centering
    \includegraphics[width=0.5\linewidth]{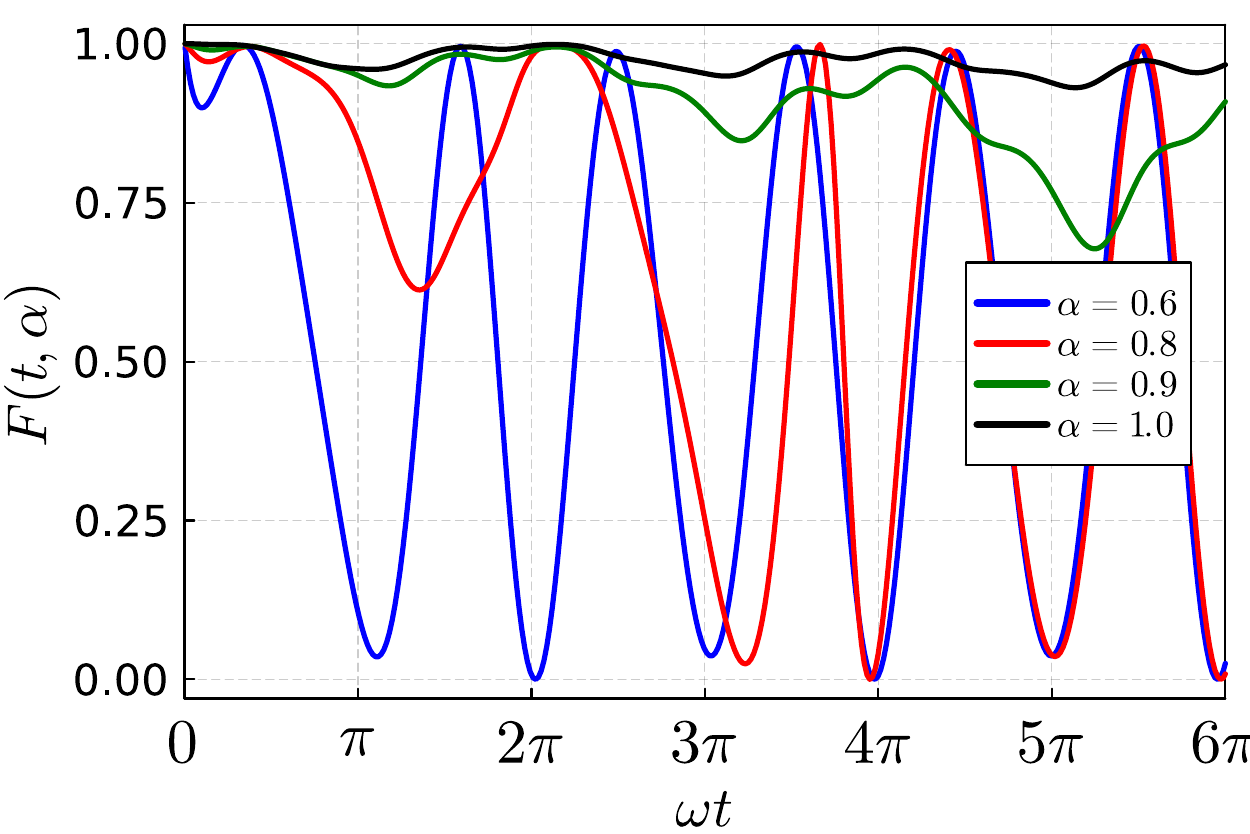}
    \caption{Time evolution of fidelity as given in equation (68), for $\Omega=\Delta/\hbar$ and $\lambda=0.1$. For the initially well separated evolutions generated by $\alpha=0.6$ and $0.8$, the memory effects decay at longer times which is a desirable property of non Markovian systems. As expected, the approximate leading order solution breaks down for $\alpha= 1$.}
    \label{fig:placeholder}
\end{figure}

On the other hand, the autocorrelation function is given by
\begin{equation}
A(t)=|c_+|^2|g_+(t)|^2+|c_-|^2|g_-(t)|^2+2\Re[c^{*}_+c_-g_+(t)g^{*}_-(t)],
\end{equation}
which upon using again the initial state given in eq. (\ref{initial-theta}),
reduces to
\begin{equation}
A(t)=\cos^2\theta/2|g_+(t)|^2+\sin^2\theta/2|g_-(t)|^2+\sin\theta\Re[g_+(t)g^{*}_-(t)],
\end{equation}
or, equivalently,
\begin{equation}
A(t)=\frac{1+(|g_+(t)|^2-|g_-(t)|^2)\cos\theta}{2}+\Re[g_+(t)g^{*}_-(t)]\sin\theta.
\end{equation}

The time evolution of these two quantities for fixed values of $\lambda=0.1$ and $\Omega=\Delta/\hbar$is shown in \figurename{3} and \figurename{4}. Interestingly, the dynamical properties of these two quantities appear to be complementary. For earlier times, the fidelity clearly shows the effects of memory for the fractional parameter values $\alpha=0.6, 0.8$, while the asymptotic behavior shows a fading of the memory effects and the curves tend to synchronize, meaning that the driving protocol stabilizes the dynamics. Thus, the transient initial regime leads to asymptotic steady oscillations in fidelity. However, the leading order description tends to break down, particularly for $\alpha=1$, where it is clearly seen that  even for small values of the effective light-matter interaction strength $\lambda=1$ the fidelity does not remain constant in time, and higher order contributions need to be included to better deal with this limiting case. For the autocorrelation function, we find that at earlier times the competition among the driving and memory effects does not show qualitative dependence on $\alpha$, but after the first quantum revival, it is manifest that lower values of the fractional order parameter dominate over the light-matter coupling and the partial revivals for $\alpha=0.6$ rapidly decay in amplitude (see line in blue colour). The intermediate value $\alpha=0.8$ shows two complete quantum revivals and one partial, meaning that the driving interaction asymptotically starts dominating over the memory (red curve), while the values $\alpha=0.9, 1$ essentially show the periodicity of the time-dependent perturbation, but, as discussed in the case of the fidelity, more terms need to be included in the perturbative solution to assert this expected physical picture.

\section{Conclusions} In this work, we have analyzed the physical consequences of fractional order quantum mechanical evolution using the Green's function approach. We have derived explicit results for the iterative calculation of the evolved  state when the dynamics is governed by the Caputo fractional derivative. Then, we have applied our results to the physically relevant two-level system, using an extension of the paradigmatic Rabi problem. We have found that even from the static contribution to the Hamiltonian one obtains non-trivial dynamic evolution of the spin polarization with damping effects related to the non local temporal evolution. Upon introduction of the driving field, the competition between energy injection due to the driving field and the nonlocal nature of the dynamical generator provide a richer physical scenario that is reflected in the spin component, autocorrelation function and fidelity evolutions in time. Although in the standard formulation, these quantities show a periodic oscillation in time, with the associated Rabi frequency measuring the rate of  oscillations of the two level system populations, within the fractional regime the memory effects take over, and these oscillations show a damping nature that can be controlled with the degree of fractionality of the time derivative. The distinctive features of the system could be experimentally tested through the observation of the fidelity or Loschmidt echo and autocorrelation function, which have been shown to dynamically evolve in a complementary fashion. Thus, the introduction of the driving interaction, to leading order in the perturbation strength, permits to achieve non-trivial evolution of the physical quantities of interest. Our results  could pave  the way to explore non-trivial dynamical effects in other systems that can also be described by means of the TLS approximation. For instance, graphene and related materials, and the SSH model for one-dimensional topological insulators, whose integer order dynamical features show the emergence of non-trivial topological phases.
\begin{figure}
    \centering
    \includegraphics[width=0.5\linewidth]{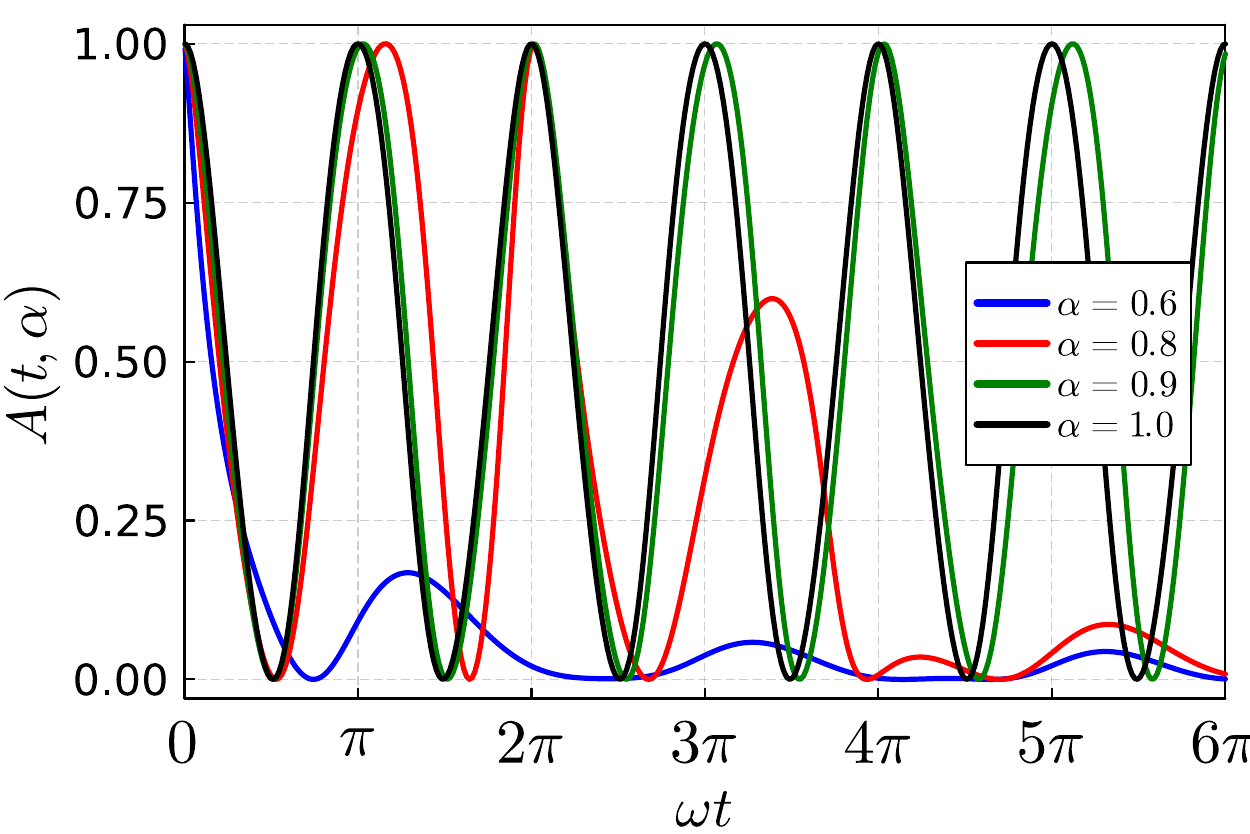}
    \caption{Time evolution of the autocorrelation, for different values of the fractional parameter $\alpha$, for $\lambda=0.1$ and $\Omega=\Delta/\hbar$. Partial quantum revivals persists up to $\alpha=0.8$ Interestingly, the closer $\alpha$ is to one, the fading of memory effects is faster. For $\alpha=0.6$ it is seen that quantum revivals are rapidly washed out by the stronger memory effects. }
    \label{fig:placeholder}
\end{figure}

This work  was partially supported by
University Lorraine, the German Science Foundation (DFG, Grant number ME 1535/12-1) and by the Alliance of International Science Organizations (Project No. ANSO-CR-PP-2022-05).

\appsection{A}{Calculation}\label{appCalculation}
In the main text we consider the integral
\begin{equation}
I_{\alpha\beta}(x,z)=\int_z^x\frac{\D y}{(x-y)^\alpha(y-z)^\beta}.    
\end{equation}
This integral can be solved analytically by means of the change of variable
\begin{equation}
u(y)=\frac{x-y}{x-z}\rightarrow y=x-u(x-z)    
\end{equation}
such that $u(z)=1$, $u(x)=0$ and $\D y=-\D u(x-z)$. Thus, the integral becomes
\begin{equation}
I_{\alpha\beta}(x,z)=\int_1^0\frac{-(x-z)\D u}{u^\alpha(x-z)^\alpha[x-u(x-z)-z]^\beta}, 
\end{equation}
which can be simplified to give
\begin{equation}
I_{\alpha\beta}(x,z)=(x-z)^{1-(\alpha+\beta)}\int_0^1 \!\D u\: u^{-\alpha}(1-u)^{-\beta}.   
\end{equation}
Upon using the definition of the Euler Beta function
\begin{equation}
B(a,b)=\int_0^1 \!\D u\: u^{a-1}(1-u)^{b-1}=\frac{\Gamma(a)\Gamma(b)}{\Gamma(a+b)},\qquad a,b>0,   
\end{equation}
we get then the result reported in the main text
\begin{equation}
I_{\alpha\beta}(x,z)=\frac{\Gamma(1-\alpha)\Gamma(1-\beta)}{\Gamma(2-(\alpha+\beta))}(x-z)^{1-(\alpha+\beta)}.   
\end{equation}
\appsection{B}{Calculation of the Fidelity}\label{AppendB}
We need to evaluate
\begin{equation}
F(t)=\frac{|\langle\Psi_1(t)\ket{\Psi_\alpha(t)}|^2}{\langle\Psi_\alpha(t)\ket{\Psi_\alpha(t)}}  
\end{equation}
where $\ket{\Psi_1(t)}$ is the exact solution to the standard Rabi problem (i.e. $\alpha=1$), whose calculation we briefly review. Let us start from the Schr\"odinger euqation
\begin{equation}\label{Rabi-integer}
\II\hbar D_t\ket{\Psi_1(t)}=H_R(t)\ket{\Psi_1(t)},   
\end{equation}
and  define $\ket{\Psi_1(t)}=P(t)\ket{\Phi(t)}$, where the time-dependent unitary transformation
is given as
\begin{equation}
P(t)=e^{-\II\Omega t(\mathbf{1}+\sigma_z)/2}=
\begin{pmatrix}
 e^{-\II\Omega t}& 0\\
0&1
\end{pmatrix}.
\end{equation}
Then we get 
\begin{equation}
\II\hbar D_t\ket{\Phi(t)}=[P^\dagger(t)H_R(t)P(t)-\II\hbar P^\dagger(t)D_tP(t)]\ket{\Phi(t)}    
\end{equation}
Performing the matrix multiplication, we get
\begin{equation}
\II\hbar D_t\ket{\Phi(t)}=\left(H_F-\frac{\hbar\Omega}{2}\right)\ket{\Phi(t)},    
\end{equation}
where we have defined
\begin{equation}
H_F=(\Delta-\hbar\Omega/2)\sigma_z+\xi\sigma_x.
\end{equation}
Therefore, the transformed state $\ket{\Phi(t)}$ evolves according to the time-independent Hamiltonian and is called a Floquet state and $H_F$ is the effective Floquet Hamiltonian. This Floquet state is written as
\begin{equation}
\ket{\Phi(t)}=e^{\II\Omega t}e^{-\II H_F t/\hbar}\ket{\Phi(0)},
\end{equation}
which in turn implies
\begin{equation}
\ket{\Psi_1(t)}=e^{-\II\Omega t\sigma_z/2}e^{-\II H_F t/\hbar}\ket{\Psi(0)},    
\end{equation}
where we have used the fact that $\ket{\Phi(0)}=\ket{\Psi_1(0)}\equiv\ket{\Psi(0)}$.
For convenience, we express the initial state on the basis of the eigenstates of $H_F$, that is, $H_F\ket{f_\pm}=\pm\epsilon\ket{f_\pm}$, which read explicitly
\begin{equation}
\ket{f_\pm}=\cos\left(\frac{\gamma}{2}+\frac{(1\mp1)\pi}{4}\right)\ket{+}+
\cos\left(\frac{\gamma}{2}-\frac{(1\pm1)\pi}{4}\right)\ket{-},
\end{equation}
where $\tan\gamma=\xi/\delta$, with the shifted energy $\delta=\Delta-\hbar\Omega/2$ associated with a one-photon resonance. Then we get
\begin{equation}\label{rabi2}
\ket{\Psi_1(t)}=e^{-\II\Omega t\sigma_z/2}(\tilde{c}_+e^{-\II \epsilon t/\hbar}\ket{f_+}+\tilde{c}_-e^{\II \epsilon t/\hbar}\ket{f_-}),    \end{equation}
where $\epsilon =\sqrt{\delta^2+\xi^2}$, the expansion coefficients are given as
\begin{eqnarray}
\tilde{c}_+&=&c_+\cos\frac{\gamma}{2}+
c_-\sin\frac{\gamma}{2}\\
\tilde{c}_-&=&c_-\cos\frac{\gamma}{2}-c_+\sin\frac{\gamma}{2}
.
\end{eqnarray}
We rewrite the leading-order solution to the fractional extension of the Rabi problem given in eq. (\ref{leading}) in the normalized form
\begin{equation}\label{leading2}
\frac{\ket{\Psi_\alpha(t)}}{\sqrt{\langle\Psi_\alpha(t)\ket{\Psi_\alpha(t)}}}\equiv\ket{\tilde{\Psi}_\alpha}=g_+(t)\ket{+}+g_-(t)\ket{-},    
\end{equation}
where 
\begin{eqnarray}\label{gpm}
g_+(t)&=&\frac{E_\alpha(-\II\omega^\alpha t^\alpha)c_++P_\alpha(t)c_-}{\sqrt{\langle\Psi_\alpha(t)\ket{\Psi_\alpha(t)}}}\nonumber\\
g_-(t)&=&\frac{E_\alpha(\II\omega^\alpha t^\alpha)c_-+P^{*}_\alpha(t)c_+}{\sqrt{\langle\Psi_\alpha(t)\ket{\Psi_\alpha(t)}}},\nonumber\\
\end{eqnarray}
where $P_\alpha(t)$ is defined in eq. (\ref{palpha}). The fidelity reads then 

\begin{equation}
F(t)=|(\tilde{c}^{*}_+e^{\II \epsilon t/\hbar}\bra{f_+}+\tilde{c}^{*}_-e^{-\II \epsilon t/\hbar}\bra{f_-})(e^{\II \Omega t/2}g_+\ket{+}+e^{-\II \Omega t/2}g_-\ket{-})|^2.
\end{equation}
 For simplicity of notation, we have  omitted the explicit time dependence in the $g_\pm$ functions.
Now we use the definition of the eigenstates of $H_F$ to rewrite this as
\begin{eqnarray}
F(t)&=&|\tilde{c}^{*}_+e^{\II \epsilon t/\hbar}(e^{\II \Omega t/2}g_+\cos\gamma/2+e^{-\II \Omega t/2}g_-\sin\gamma/2)+\tilde{c}^{*}_-e^{-\II \epsilon t/\hbar}(e^{-\II \Omega t/2}g_-\cos\gamma/2-e^{\II \Omega t/2}g_+\sin\gamma/2)|^2,\nonumber\\
\end{eqnarray}
which can also be expressed as
\begin{eqnarray}
F(t)&=&|\tilde{c}_+|^2( |g_+|^2\cos^2\gamma/2+|g_-|^2\sin^2\gamma/2+\Re(g_+g^{*}_-e^{\II \Omega t} )\sin\gamma+\nonumber\\
&&|\tilde{c}_-|^2( |g_-|^2\cos^2\gamma/2+|g_+|^2\sin^2\gamma/2- \Re(g_+g^{*}_-e^{\II \Omega t})\sin\gamma+\nonumber\\
&&2\Re[\tilde{c}^{*}_+\tilde{c}_-e^{2\II \epsilon t/\hbar}(e^{\II \Omega t/2}g_+\cos\gamma/2+e^{-\II \Omega t/2}g_-\sin\gamma/2)(e^{\II \Omega t/2}g^{*}_-\cos\gamma/2-e^{-\II \Omega t/2}g^{*}_+\sin\gamma/2)],\nonumber\\
\end{eqnarray}
which is equivalent to
\begin{eqnarray}
F(t)&=&\frac{|\tilde{c}_+|^2[|g_+|^2+|g_-|^2+(|g_+|^2-|g_-|^2)\cos\gamma+2\Re(g_+g^{*}_-e^{\II \Omega t} )\sin\gamma]}{2}+\nonumber\\
&&\frac{|\tilde{c}_-|^2[|g_+|^2+|g_-|^2-(|g_+|^2-|g_-|^2)\cos\gamma-2\Re(g_+g^{*}_-e^{\II \Omega t} )\sin\gamma]}{2}+\nonumber\\
&&2\Re\left[\tilde{c}^{*}_+\tilde{c}_-e^{2\II \epsilon t/\hbar}\left(e^{\II \Omega t}g_+g^{*}_-\cos^2\gamma/2-e^{-\II \Omega t}g_-g^{*}_+\sin^2\gamma/2
-\frac{(|g_+|^2-|g_-|^2)}{2}\sin\gamma\right)\right],\nonumber\\
\end{eqnarray}
Collecting similar terms together and using the normalization conditions $|g_+|^2+|g_-|^2=1$ and $|\tilde{c}_+|^2+|\tilde{c}_-|^2=1$, the previous expression is further reduced to take the form:
\begin{eqnarray}
F(t)&=&\frac{
 1+(|\tilde{c}_+|^2-|\tilde{c}_-|^2)\left[\left(|g_+|^2-|g_-|^2\right)\cos\gamma+2\Re(g_+g^{*}_-e^{\II \Omega t})\sin\gamma\right]}{2}+\nonumber\\
&& 2\Re\left\{\tilde{c}^{*}_+\tilde{c}_-e^{2\II \epsilon t/\hbar}\left[\II\Im (e^{\II \Omega t}g_+g^{*}_-)+\Re(e^{\II \Omega t}g_+g^{*}_-)\cos\gamma
-\frac{(|g_+|^2-|g_-|^2)}{2}\sin\gamma\right]\right\},\nonumber\\
\end{eqnarray}
which in turn is further simplified to read
\begin{eqnarray}
F(t)&=&\frac{
 1+(|\tilde{c}_+|^2-|\tilde{c}_-|^2)[\left(|g_+|^2-|g_-|^2\right)\cos\gamma+2\Re(g_+g^{*}_-e^{\II \Omega t})\sin\gamma]}{2}+\nonumber\\
&& -2\Im(\tilde{c}^{*}_+\tilde{c}_-e^{2\II \epsilon t/\hbar})\Im (e^{\II \Omega t}g_+g^{*}_-)+2\Re(\tilde{c}^{*}_+\tilde{c}_-e^{2\II \epsilon t/\hbar})\left[\Re(e^{\II \Omega t}g_+g^{*}_-)\cos\gamma
-\frac{(|g_+|^2-|g_-|^2)}{2}\sin\gamma\right].\nonumber\\
\end{eqnarray}
We can check $F(0)=1$. For that purpose, we recall that $g_\pm(0)=c_\pm$, thus
\begin{eqnarray}
F(0)&=&\frac{
 1+(|\tilde{c}_+|^2-|\tilde{c}_-|^2)[\left(|c_+|^2-|c_-|^2\right)\cos\gamma+2\Re(c_+c^{*}_-)\sin\gamma]}{2}+\nonumber\\
&& -2\Im(\tilde{c}^{*}_+\tilde{c}_-)\Im (c_+c^{*}_-)+2\Re(\tilde{c}^{*}_+\tilde{c}_-)\left[\Re(c_+c^{*}_-)\cos\gamma
-\frac{(|c_+|^2-|c_-|^2)}{2}\sin\gamma\right].\nonumber\\
\end{eqnarray}
Using the definition of $\tilde{c}_\pm$, we get
\begin{eqnarray}
|\tilde{c}_+|^2-|\tilde{c}_+|^2&=& \left(|c_+|^2-|c_-|^2\right)\cos\gamma+2\Re(c_+c^{*}_-)\sin\gamma\nonumber\\
\Re(\tilde{c}^{*}_+\tilde{c}_-)&=&\Re(c_+c^{*}_-)\cos\gamma
-\frac{(|c_+|^2-|c_-|^2)}{2}\sin\gamma\nonumber\\
\Im(\tilde{c}^{*}_+\tilde{c}_-)&=&\Im(c^{*}_+c_-),\nonumber
\end{eqnarray}
which leads to
\begin{eqnarray}
F(0)&=&\frac{
 1+(|\tilde{c}_+|^2-|\tilde{c}_-|^2)^2}{2}+2[\Im(\tilde{c}^{*}_+\tilde{c}_-)]^2+2[\Re(\tilde{c}^{*}_+\tilde{c}_-)]^2.\nonumber\\
\end{eqnarray}
Using the normalization condition, this can be further reduced to
\begin{eqnarray}
F(0)&=&
1-2|\tilde{c}_+|^2|\tilde{c}_-|^2+2[\Im(\tilde{c}^{*}_+\tilde{c}_-)]^2+2[\Re(\tilde{c}^{*}_+\tilde{c}_-)]^2=1,\nonumber\\
\end{eqnarray}
as desired.
\subsection{Resonant regime}
Interestingly, at resonance $\Omega=2\omega$, we obtain $\gamma=\pi/2$, which also implies $\epsilon_{res}=\lambda\hbar\omega$. Therefore, the resonant evolution for the fidelity reads:
\begin{eqnarray}
F_{res}(t)&=&\frac{
1+2\Re(g_+g^{*}_-e^{\II \Omega t})\sin\theta}{2}+ \cos\theta\left[\sin{(\lambda\omega t})\Im (e^{\II \Omega t}g_+g^{*}_-)+\cos(\lambda \omega t)\frac{(|g_+|^2-|g_-|^2)}{2}\right].\nonumber\\
\end{eqnarray}
For small values of $\lambda\ll1$, consistent with the leading-order  fractional dynamical behavior of interest, we can further approximate
\begin{eqnarray}
F_{res}(t)&\approx&\frac{
1+2\Re(g_+(t)g^{*}_-(t)e^{\II \Omega t})\sin\theta}{2}+ \left(\frac{|g_+(t)|^2-|g_-(t)|^2}{2}\right)\cos\theta.\nonumber\\
\end{eqnarray}
This resonant expression could be used in experimental setups to further gain insight on the interplay of periodic driving and the memory effects due to the fractional dynamical generator. 

\end{document}